# Dynamic Channel Allocation for Class-Based QoS Provisioning and Call Admission in Visible Light Communication

Mostafa Zaman Chowdhury, Muhammad Shahin Uddin, and Yeong Min Jang

*Department of Electronics Engineering, Kookmin University, Seoul 136-702, Korea.*

E-mail: mzceee@yahoo.com, shahin.mbstu@gmail.com, yjang@kookmin.ac.kr

**Abstract**  Provisioning of quality of service (QoS) is a key issue in visible light communication (VLC) system as well as in other wireless communication systems. Due to the fact that QoS requirements are not as strict for all traffic types, more calls of higher priority traffic classes can be accommodated by blocking some more calls of lower priority traffic classes. Diverse types of high data rate traffic are supported by existing wireless communication systems while the resource is limited. Hence, priority based resource allocation can ensure the service quality for the calls of important traffic class. The fixed guard channels to prioritize any class of calls always reduce the channel utilization. In this paper we propose a priority based dynamic channel reservation scheme for higher priority calls that does not reduce the channel utilization significantly. The number of reserved channels for each of the individual traffic classes is calculated using real-time observation of the call arrival rates of all the traffic classes. The features of the scheme allow reduction of the call blocking probability of higher priority calls along with the increase of the channel utilization. The proposed Markov Chain model is expected to be very much effective for the queuing analysis especially for the priority scheme of any number of traffic classes. The numerical results show that the proposed scheme is able to attain reasonable call blocking probability of higher priority calls without sacrificing channel utilization.

**Keywords** *QoS, visible light communication, priority users, call blocking probability, and channel allocation.*

## 1. Introduction

The trend of wireless communication systems is the increase in the variety of multimedia applications, which further spreads the traffic load of wireless networks. Among the variety of traffic, few classes of traffic e. g., traffic related to security, healthcare, banking, handover calls, and etc. are more important than others. During the resource allocation, these important classes of traffic are given higher priority through different mechanisms. It is preferable that, blocking a low priority call than blocking of the higher priority calls when the system's resources are running low. Several prior research works (e.g., in [1]-[5]) have been done that allow higher priority for a particular class of traffic (e.g. handover call) over others. Most of these proposed schemes are based on the notion of "fixed guard band" and "quality of service (QoS) adaptability". In "fixed guard band," a number of channels are reserved for the exclusive use of a particular class of traffic. However, the schemes based on "QoS adaptability" reduce the resource allocation for the running low priority traffic calls to accept more calls of higher priority traffic in the system. Although the schemes based on fixed guard bands [1] are simple and reduce the call blocking probability of higher priority traffic calls, these schemes also result in reduced bandwidth utilization. The ability of QoS adaptability [2]-[5] of the traffic allows to give higher priority for the higher priority traffic classes over the lower classes. However, these schemes have the limitations that, the system have to permit variable bandwidth allocation for the calls (e.g., the circuit switching system does not allow the fraction of channel allocation) as well as the service applications must have the ability of QoS adaptability (e.g., the real-time voice calls need guaranteed bandwidth). Therefore, a dynamic channel allocation scheme is required to provide the priority based resource allocation for the non-QoS adaptive communication systems.

Recently, the green and energy-efficient wireless communication systems have experienced rapid development and garnered much interest. One such scheme, the visible light communication (VLC) system in which visible light is used as a high data rate transmission medium. It is being touted as one of the most promising wireless communication technologies for the next generation. VLC is a short-range optical wireless communication utilizing LED lighting, so that the LED lights can provide both illumination and communication simultaneously. Nowadays, there are a growing number of scenarios; in house domestic applications, traffic lights, hospitals, hotels, etc. where illumination devices based on visible LEDs are used. These LEDs combine very low power consumption with an extremely long operational life, small size, and rapid response time. The LEDs also maintain same chromaticity without significant changes during all their operations. On the other hand, LEDs can be used as communication transmitter without losing their main functionality as illumination sources. The installation of a wireless network using visible light based on an existing interior LEDs lighting infrastructure would probably be easier and cost-effective. Visible light communications not only maintain the usual capabilities of wireless optical transceivers but also have robustness against electromagnetic (EM) interference and security against undesired network access. From a security perspective, VLC is slightly different from other wireless networks due to directionality and visibility of the visible optical spectrum. As a result of directionality and visibility, if an unauthorized receiver is in the path of the communication signal, it can be recognized. Also, the signal does not travel across medium such as walls, unlike other radio frequency based wireless networks. These features of VLC make this system suitable for co-existing with commercial RF networks WiFi, Bluetooth, etc., especially for in-house applications.

The resource management is an important module for the VLC system to guarantee the QoS for diverse type of users. In this paper, we propose a priority based channel allocation for VLC. In VLC system, each channel has its own color according to the draft VLC specification frequency band plan [6]. If multi-class of traffic such as voice, video, data, etc. are aggregated in the system, then each of the calls of different traffic classes can use different color channel. Resource is allocated for each type of users based on the priority and instantaneous color channel condition for maintaining the QoS in the VLC system. The priority of all the aggregated traffic classes is not same to meet QoS requirements. For the VLC networks, due to high data rate services and limited number of available channels, only an efficient priority based call admission control (CAC) mechanism can ensure very low call blocking probability during heavy traffic condition for the higher priority users e.g., handover calls, link recovery [7] calls, conversational voice calls, and etc. without sacrificing resource utilization. This is because, the non-priority scheme where all the traffic calls are treated equally, results very high call blocking probability for the higher priority traffic calls during the higher traffic condition. The proposed priority based non-fixed reserved channel scheme increases the channel utilization as well as reduces the call blocking probability of the higher priority calls. Call connection requests are classified into multi-classes such as class 1, class 2, … … class *M*. The number of available channels to accept a call for a particular class of traffic is varied with the variation of call arriving rates of all the traffic classes. The non-fixed numbers of guard channels maintain the higher channel utilization along with the lower call blocking probability of higher priority traffic calls. The real-time traffic arrival rate estimation is applied for the proposed scheme to calculate the number of reserved channels for each of the traffic classes.

The rest of this paper is organized as follows. Section 2 introduces the service scenario and optical channel model for the visible light communication. Proposed dynamic channel allocation scheme including the system model and queuing analysis are shown in Section 3. We also derive the formulas for the call blocking probability for any class of traffic calls in this Section. Section 4 presents the numerical performance evaluation results of our proposed schemes. Finally, we conclude our work in Section 5.

## 2. VLC Service Scenario and Optical Channel Model

### 2.1 Service Scenario

A basic indoor VLC service scenario is shown in Fig. 1. Lighting and communication sources are provided by a number of LED arrays disposed on the room ceiling. A terminal with optical receiver is placed on a receiving plane, such as a desk, and the data is received from the illuminating sources. In Fig.1, the VLC network is connected to the power line communication (PLC) network or cellular network or IP network. VLC network is used for the purposes of high-definition television (HDTV) services, audio services, high speed internet services, video streaming services, different types of smart phone users, and etc. The diverse types of traffic require different level of QoS for each of them. Based on the priorities of the traffic classes and their requirements, channels are allocated dynamically for each of the traffic classes to meet their desired QoS levels.

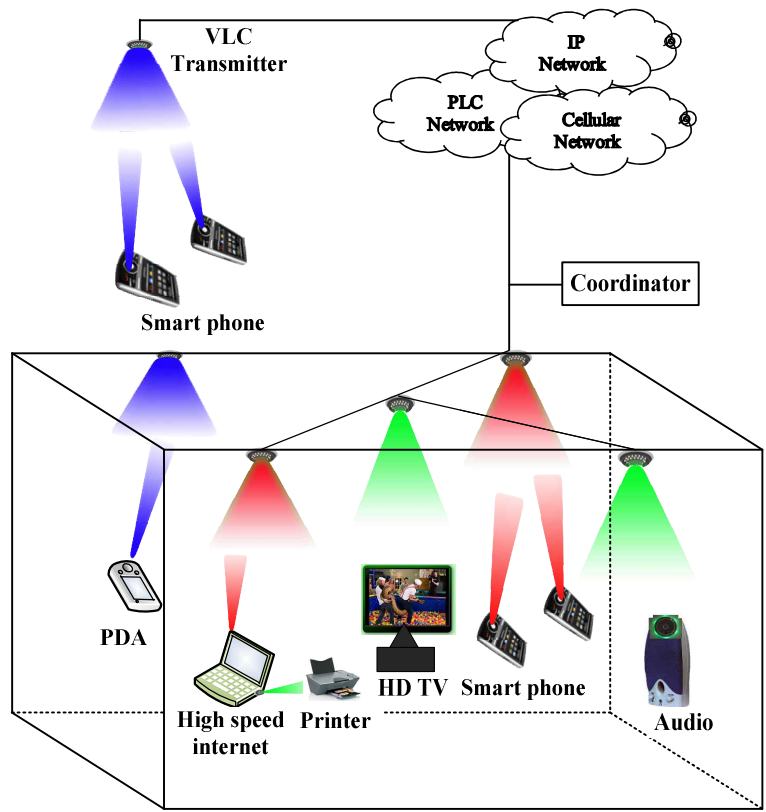

**Fig. 1.** A basic network architecture and indoor service scenario for visible light communication.

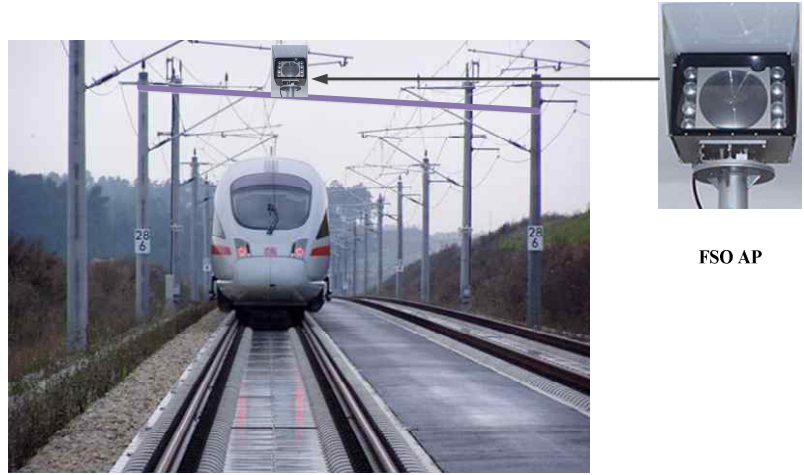

**Fig. 2.** VLC outdoor application scenario where FSO communication is used for the users in train.

Fig. 2 shows one example of VLC in outdoor environment. Free-space optical communication (FSO), one kind of visible light communication is an optical communication technology that uses light propagating in free space to transmit data for telecommunications. In the highway road or train line or subway line, the FSO communication networks can be installed. The FSO access point (AP) are installed for every 2 km or 3 km. The FSO AP are connected to the optical fiber backbone. Optical transceivers are installed in outside a vehicle. Inside the vehicle, WLAN, femtocell [8], or other indoor wireless networks are installed. The users are connected to these networks. However, for backhauling the traffic for these networks, the FSO network is used. This application of VLC can support huge number of users in vehicles.

In VLC system, line-of-sight (LOS) link between two transceivers should be guaranteed due to the straightness of the visible light signal. Also, visibility support is needed in the VLC networks to make the easy initial access, for the link re-connection, and to identify the nature of the obstacle that interrupts the communication. This visibility support requires the use of some optical channels

## 2.2 Optical Channel Model and Color Band

In VLC service scenario, each class of users use the visible light link to receive or transmit the data. For high data rate applications, link should be line-of-sight. The bandwidth of the optical channel in LOS configuration is reported higher than 88 MHz [9]. Here, we are going to show the behavior of the optical channel when the visible optical signal is passing from transmitter to the receiver. The received power depends on the optical channel gain and the transmitted power. The optical channel gain that is related to transmitted and received powers can be expressed as:

$$P_r = H(0)P_t \qquad (1)$$

where $P_t$ is the transmitted optical power, $P_r$ is the received optical power, and $H(0)$ is the channel DC gain.

Considering the LOS link, the channel DC gain is defined as [10]:

$$H_{LOS} = \begin{cases} \dfrac{(\tau+1)A}{2\pi D^2} \cos^\tau(\varphi) T_S(\psi) g(\psi) \cos(\psi), & 0 \leq \psi \leq \psi_c \\ 0, & elsewhere \end{cases} \quad (2)$$

where $\tau$ is the order of Lambertian emission, $A$ is the photo-detector area, $D$ is the distance between the transmitter and receiver, $\varphi$ is the angle of irradiance, $\psi$ is the angle of incidence, $T_s(\psi)$ is the signal transmission coefficient of an optical filter, $g(\psi)$ is the gain of an optical concentrator, and $\psi_c$ is the receiver field of view (FOV).

The order of Lambertian emission $\tau$ can be found from the equation, $\tau = -\dfrac{\ln 2}{\ln(\cos \phi_{1/2})}$, where $\phi_{1/2}$ is the transmitter half power angle. The gain can be determined from the following expression [10]:

$$g(\psi) = \begin{cases} \dfrac{v^2}{\sin^2(\psi_c)}, & 0 \leq \psi \leq \psi_c \\ 0, & elsewhere \end{cases} \quad (3)$$

where $v$ denotes the internal refractive index of the optical concentrator.

VLC device operates in one or several color bands with peak radiated energy within the visible light wavelength spectrum from 380 nm to 780 nm as summarized in Table 1.

**Table 1** Visible light color band plan

| Color band | Wavelength (nm) | Spectral width (nm) |
|:---:|:---:|:---:|
| Band 1 | 380 - 450 | 70 |
| Band 2 | 450 - 510 | 60 |
| Band 3 | 510 - 560 | 50 |
| Band 4 | 560 - 600 | 40 |
| Band 5 | 600 - 650 | 50 |
| Band 6 | 650 - 710 | 60 |
| Band 7 | 710 - 780 | 70 |

Human eye sensitivity is not same for all the color bands. Therefore, visible LEDs are designed to match human eye sensitivity and to support up to 7 independent and parallel bands according to the VLC draft specification [6]. Using the combination of these seven bands we can make maximum 27-1 multiple channels (0000000 is not used). Therefore, the maximum number of channel is limited in VLC system. Bit patterns for single or multiple color bands selection are summarized in Table 2 [6]. Bit patterns 0000001 and 1000000, respectively, represent the selected color band 7 and color band 1. Bit pattern 0100110 represents that the selected bands for communication are 2, 5, and 6. In this way 1111111 represents that all the 7 color bands are selected for communication. When VLC device supports multiple color bands, one can optimize the link to choose the desired band for best performance and network capacity. If a VLC device transmits on a certain color band, then it may not want to receive on the same color band. In this case, the device can use another color band for receiving. Knowing what color bands are being used for transmission and what color bands will support, can influence receiver's choice for reverse link transmission.

**Table 2** Color band selection for multiple channel (*X=not used and O=used*)

| Bit Pattern | Band 1 | Band 2 | Band 3 | Band 4 | Band 5 | Band 6 | Band 7 |
|---|---|---|---|---|---|---|---|
| 0000001 | X | X | X | X | X | X | O |
| 0000010 | X | X | X | X | X | O | X |
| … | | | | | | | |
| 0100110 | X | O | X | X | O | O | X |
| … | | | | | | | |
| 1000000 | O | X | X | X | X | X | X |
| … | | | | | | | |
| 1111111 | O | O | O | O | O | O | O |

## 3. Proposed Channel Allocation Scheme

The radio resource management module is responsible for the efficient utilization of air interface resources to guarantee a certain QoS level for different users according to their traffic profiles [11]. Many priority schemes for mobile networks have been already proposed by different researchers to reduce the handover call dropping probability or to reduce the call blocking probability of higher priority calls [1]-[5], [12], [13]. If fixed guard channels [1] are used to prioritize any class of calls then the system resource utilization is always reduced. Also, if the scheme is based on the QoS adaptability [2]-[5] then the system and the application have to be cope with the QoS adaptability that may not possible for every wireless networks and service environments. The channel borrowing scheme [5] results increased signaling overhead due to communication with the neighboring cells. Therefore, we need a priority scheme for the non-QoS adaptive environments (the non-QoS adaptive calls do not allow the reduction of bandwidth allocation for them) to give higher priority for the important class of traffic calls. A priority based dynamic channel allocation scheme for the VLC networks ensures the lower call blocking probability of the higher priority users and maintains higher channel utilization. In this section, we propose a non-fixed guard channel scheme that increases the channel utilization as well as reduces the call blocking probability of the higher priority calls. The number of available channels to accept a call for a particular class of traffic is varied with the variation of call arriving rates of all the traffic classes. We propose the queuing analysis model to calculate the call blocking probability and technique to estimate the traffic arrival rate. Even though, we use VLC as a model for the proposed scheme, the proposed scheme is applicable for any wireless networks especially those do not support the QoS adaptability.

### 3.1 Dynamic Channel Allocation

Let, traffic classes *1, 2, 3, ..., m, ..., M,* respectively, represent the highest to lowest priority. $\lambda_T$ and $\lambda_m$ are, respectively, the total call arriving rate and the call arriving rate of traffic class *m*. So,

$$\lambda_T = \lambda_1 + \lambda_2 + \ldots + \lambda_m + \ldots + \lambda_M \tag{4}$$

Suppose, $N$ is the total number of channels in the system and $\Gamma$ is the maximum allowable number of channels only those are used for the reservation purposes (equivalently guard channels). The reservation of channels for various traffic classes are based on the priority of the traffic classes. The remaining *(N-Γ)* number of channels are equally shared by all the traffic classes. The value of Γ should not be very high in order to maintain the higher channel utilization.

The number of reserved channels for only the traffic classes of 1, 2, …, and *m* is:

$$X_m(t) = \frac{\lambda_m}{\lambda_T} \Gamma \tag{5}$$

Alternatively, the term $\Gamma$ can be expressed as:

$$\Gamma = \sum_{i=1}^{M} X_i(t) \tag{6}$$

Along with *(N- Γ)* channels, $\lfloor X_M(t) \rfloor$ number of channels are also equally shared by all the traffic classes.

The number of accessible channels among $\Gamma$, for the *m-th* class traffic calls is:

$$y_m(t) = \left\lfloor \sum_{i=m}^{M} X_i(t) \right\rfloor \tag{7}$$

Total number of accessible channels for the traffic class *m* is:

$$N_m = N - \Gamma + y_m(t) \tag{8}$$

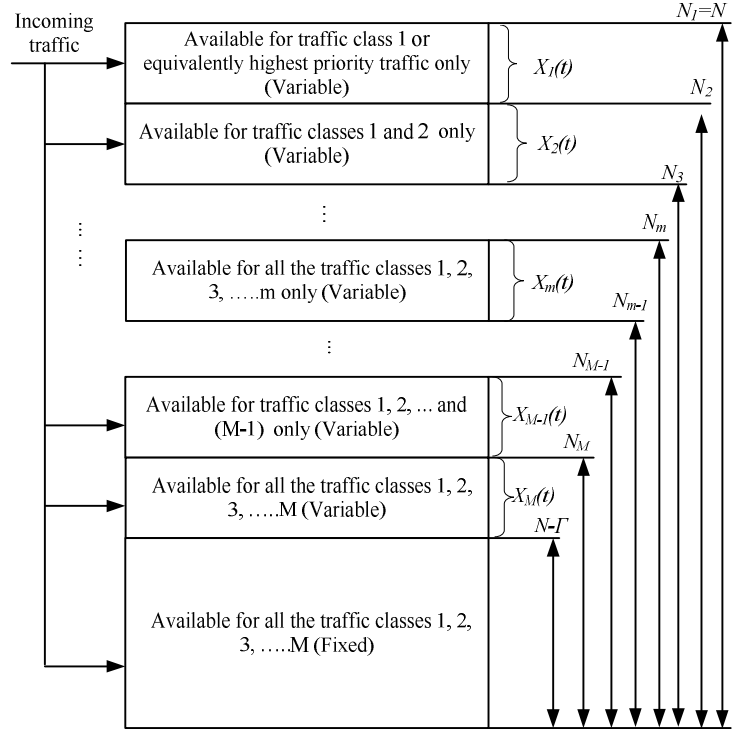

**Fig. 3.** System model for the proposed scheme to admit a call.

Equations (5) - (8) show that the channel allocations for the different traffic classes are not fixed. Depending on $\lambda_m$, $\lambda_T$, $N$, and $\Gamma$, the numbers of channel allocations are varied. At a particular time, the number of maximum channels $N_m$ to accept a call of the traffic class $m$ can be calculated by measuring the traffic arrival rates.

Fig. 3 shows the system model for the proposed dynamic channel allocation scheme. The proposed scheme reserves a non-fixed number of channels for the traffic class $m$. The number of available channels depends on the priority of the traffic class and the call arriving rates of different traffic classes. ($N$-$\Gamma$) number of channels are available for all the traffic classes, whereas the remaining $\Gamma$ number of channels are allocated based on the priority of the traffic classes. If the priority of a traffic class is higher, then comparatively higher number of channels are available for that traffic class. A call request of traffic class $m$ can be accepted only if the number of already occupied channels by the existing calls is less than $N_m$.

### 3.2 Queuing Analysis

The CAC for the proposed scheme is shown in Fig. 4. Whenever a call of traffic class $m$ is arrived, the system estimates the traffic arrival rates of all the traffic classes to determine the maximum number of accessible channels $N_m$ for the traffic class $m$. A call of traffic class $m$ can be accepted only if the number of already occupied channels by the existing calls is less than $N_m$.

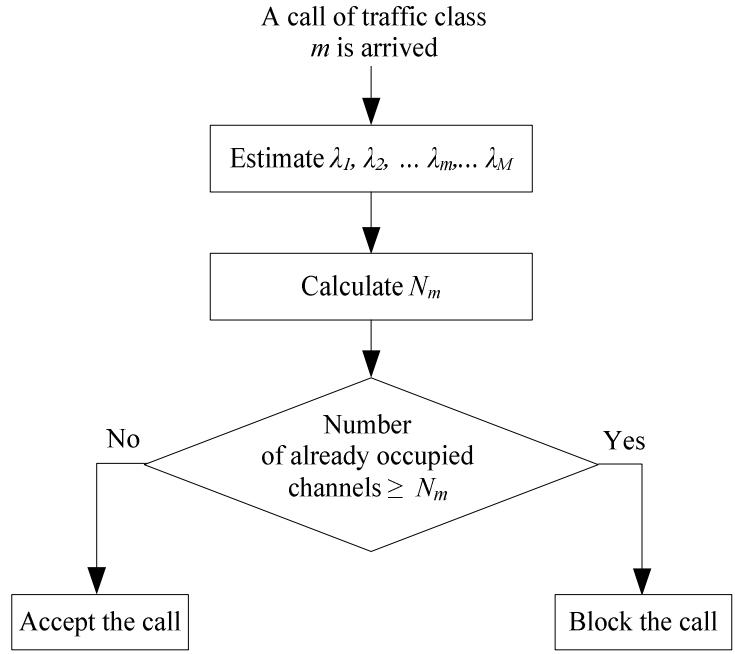

**Fig. 4.** A call admission control (CAC) for the proposed scheme.

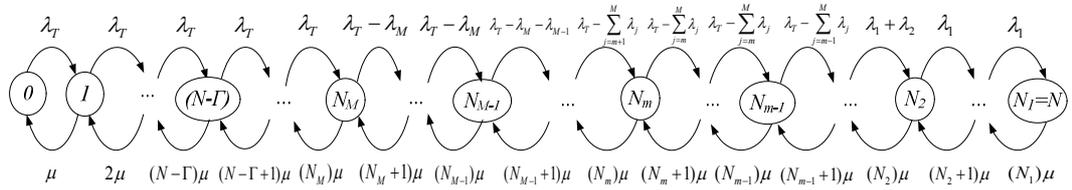

**Fig. 5.** Markov Chain for the queuing analysis of the proposed scheme.

The proposed scheme can be modeled as an *M/M/N/N* queuing system [14]. The Markov Chain for the queuing analysis of the proposed scheme is shown in Fig. 5. We define $1/\mu$ as the average channel holding time (exponentially distributed). The probability that the system is in state *i*, is given by $P_i$. From the Fig. 5, the state balance equations are expressed as:

$$\begin{cases} i\mu P_i = \lambda_T P_{i-1}, & 0 \leq i \leq N_M \\ i\mu P_i = (\lambda_T - \lambda_M)P_{i-1}, & N_M < i \leq N_{M-1} \\ i\mu P_i = (\lambda_T - \sum_{j=m}^{M}\lambda_j)P_{i-1}, & N_m < i \leq N_{m-1} \\ i\mu P_i = (\lambda_1 + \lambda_2)P_{i-1}, & N_3 < i \leq N_2 \\ i\mu P_i = \lambda_1 P_{i-1}, & N_2 < i \leq N_1 \end{cases} \quad (9)$$

A call of *m-th* class traffic is blocked in the proposed scheme if the state of the system calls is $N_m$ or larger. $N_M$ is the maximum number of available channels for the lowest priority (*M-th* class) traffic class whereas $N_1$ is the maximum number of available channels for the highest priority (traffic class 1) traffic class. The highest priority calls are blocked only when all the total *N* number of channels (also denoted by $N_1$) are occupied by the existing calls. The generalized equations to calculate the blocking probability for any call of traffic class *m* among total *M* number of traffic classes are derived using the queuing analysis. The call blocking probability for a call of highest priority traffic class is calculated using (10). Whereas the call blocking probability for the traffic classes from 2 and higher are calculated using (11). This Markov Chain model can be effectively applied for the queuing analysis especially for the priority scheme of any number of traffic classes.

$$B_1 = P_N = \frac{\lambda_T^{N_M}}{\mu^N N!} P_0 \prod_{k=1}^{M-1} (\lambda_1 + \lambda_2 + \ldots + \lambda_{M-k})^{N_{M-k} - N_{M-k+1}}, \quad m=1 \quad (10)$$

$$B_m = \sum_{i=N_m}^{N} P_i$$
$$= B_{m-1} + P_0 \sum_{i=N_m}^{N_{m-1}-1} \frac{\lambda_T^{N_M}}{\mu^i i!} (\lambda_1 + \lambda_2 + \ldots + \lambda_{m-1})^{i-N_m} \prod_{k=m}^{M-1} (\lambda_1 + \lambda_2 + \ldots + \lambda_k)^{N_k - N_{k+1}}, \quad 2 \le m \le M \quad (11)$$

$$P_0 = \left[ 1 + \sum_{i=1}^{N_M} \frac{\lambda_T^i}{\mu^i i!} + \sum_{j=2}^{M} \sum_{i=N_j+1}^{N_{j-1}} \left\{ \frac{\lambda_T^{N_M} (\lambda_1 + \lambda_2 + \ldots + \lambda_{j-1})^{i-N_j}}{\mu^i i!} \prod_{k=j}^{M-1} (\lambda_1 + \lambda_2 + \ldots + \lambda_k)^{N_k - N_{k+1}} \right\} \right]^{-1} \quad (12)$$

### 3.3 Estimation of Traffic Arrival Rate

For the estimation of the traffic call arrival rate, last *(n+1)* number of arriving calls for every traffic classes are observed. The time interval $\Delta t_i^m$, between the *(i-1)-th* and *i-th* call arrivals of *m-th* class traffic is measured. Thus, total *n* number of samples are taken for each of the traffic classes to calculate the average time interval between two successive call arrivals. Fig. 6 shows the *(n+1)* number of call arrivals of traffic class *m* for the estimation of traffic call arrival rate. $t_0^m$ refers the most recent call arrival instant of traffic class *m*.

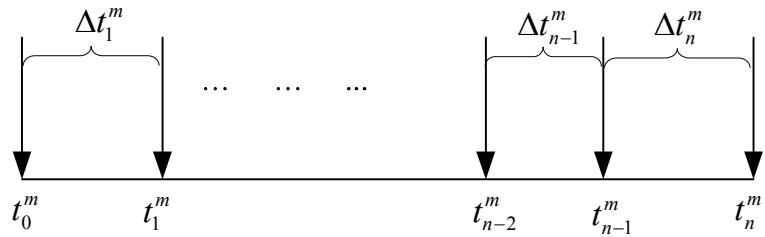

**Fig. 6.** Call arriving window that is used for the estimation of call arrival rate of traffic class *m*.

The average time interval between the two successive calls of last *(n+1)* number of call arrivals for the *m-th* class traffic is calculated as:

$$\overline{\Delta t^m} = \frac{1}{n}\sum_{i=1}^{n}\Delta t_i^m \qquad (13)$$

Now taking the expectation with respect to $\overline{\Delta t^m}$, we have

$$E[\overline{\Delta t^m}] = E\left[\frac{1}{n}\sum_{i=1}^{n}\Delta t_i^m\right] = \Delta t^m \qquad (14)$$

where $\Delta t^m$ is the true value. This shows that $\overline{\Delta t^m}$ is an unbiased estimation [15].

The average call arrival rate ($\lambda_m$) of *m-th* class traffic considering the last *(n+1)* number of call arrivals is calculated as:

$$\frac{1}{\lambda_m} = \overline{\Delta t^m} = \frac{1}{n}\sum_{i=1}^{n}\Delta t_i^m \qquad (15)$$

As, $\overline{\Delta t^m}$ is an unbiased estimation, $\lambda_m$ is also an unbiased estimation. Hence, $\lambda_m$ in (15) is used to estimate the call arrival rate of traffic class *m*. Using (5) - (7) and (13) – (15), (8) can be expressed as:

$$N_m = \left\lfloor N - \frac{\sum_{j=1}^{m-1}\frac{1}{\sum_{i=1}^{n}\Delta t_i^j}}{\sum_{j=1}^{M}\frac{1}{\sum_{i=1}^{n}\Delta t_i^j}}\Gamma \right\rfloor \qquad (16)$$

Here, we consider only last *(n+1)* number of call arrivals to emphasis the near present traffic condition. As we consider only *(n+1)* number of call arrivals for the calculation of $N_m$, whenever a latest call is arrived, the sample for the *n-th* call is replaced by the sample of the *(n+1)-th* call. Equation (16) indicates that the number of reserved channels for the traffic class *m* is high if *m* represents higher priority traffic class. Hence, the number of allocated channels for a traffic class is dynamically varied depending on the call arrival rates, *N*, and *Γ*.

## 4. Performance Evaluation

In this section, we present the results of the numerical analysis of the proposed scheme. The call arriving process is assumed to be Poisson and that the channel holding time is exponentially distributed with the average of 120 sec. We assume total 100 number of available channels in the system. $\Gamma$ is considered to be 10. We consider total four classes of traffic for our analysis. Handover (or equivalently link switched) calls, link recovery calls, and conversational voice calls are considered as traffic class 1 (priority 1); conversational video calls are considered as traffic class 2 (priority 2); internet browsing, buffered streaming video, and voice messaging are considered as traffic class 3 (priority 3); and background data traffic are considered as traffic class 4 (priority 4) for our analysis. 100 number of samples are used for each of the traffic classes to estimate the call arrival rate.

We performed call blocking probability, channel utilization, and the number of reserved channel analysis through two sets of results for different traffic arrival conditions. Figs. 7 and 8, respectively, show the comparison of call blocking probability and channel utilization performances for the condition when equal traffic arrival rates of all the four classes (1:1:1:1). Figs. 9 and 10, respectively, show the comparison of call blocking probability and channel utilization performances for the condition when traffic arrival rates of higher priority classes are comparatively higher than the lower traffic classes (e.g., 3:4:2:1).

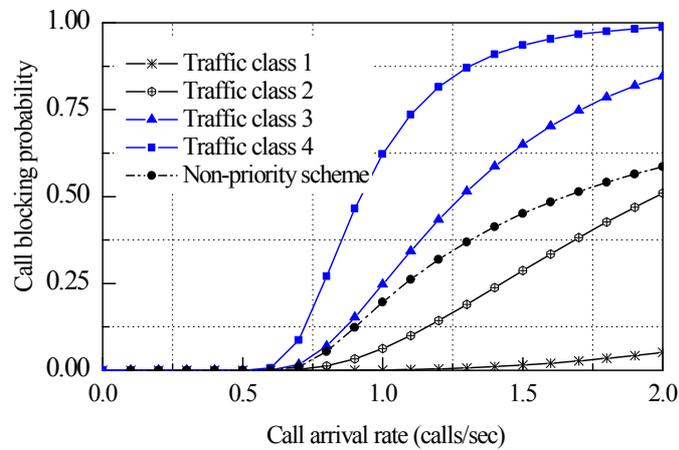

**Fig. 7.** Comparison of call blocking probability when the ratio of traffic arrival rate of traffic class 1, 2, 3, and 4 is assumed to be 1:1:1:1.

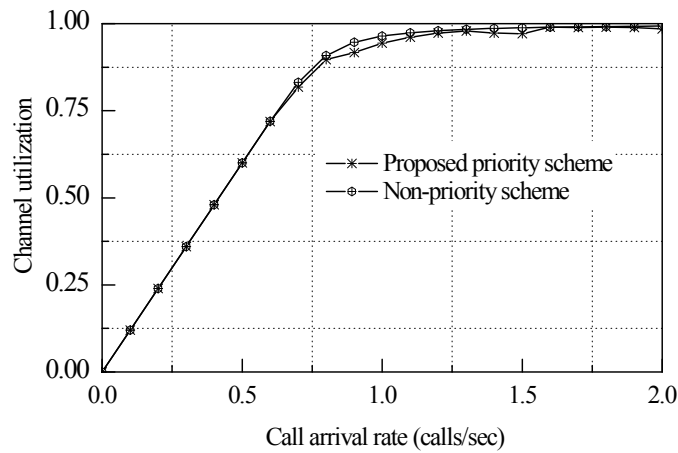

**Fig. 8.** Comparison of channel utilization when the ratio of traffic arrival rate of traffic class 1, 2, 3, and 4 is assumed to be 1:1:1:1.

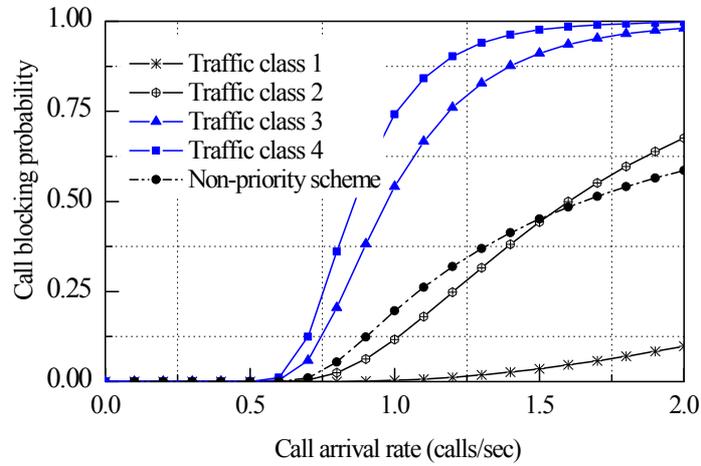

**Fig. 9.** Comparison of call blocking probability when the ratio of traffic arrival rate of traffic class 1, 2, 3, and 4 is assumed to be 3:4:2:1.

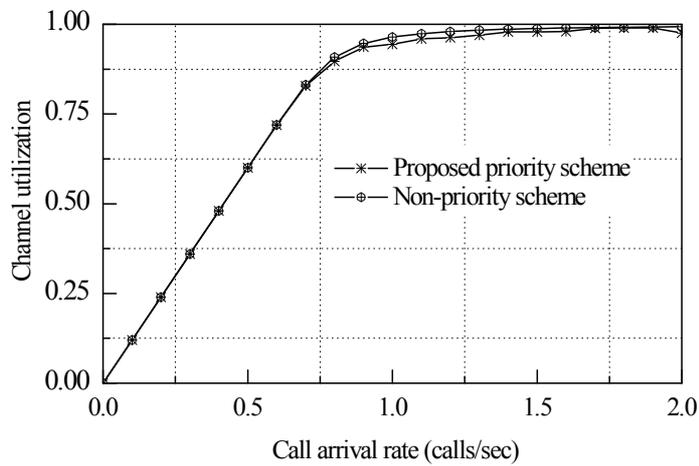

**Fig. 10.** Comparison of channel utilization when the ratio of traffic arrival rate of traffic class 1, 2, 3, and 4 is assumed to be 3:4:2:1.

The The non-priority scheme results very high call blocking probability for the higher priority traffic calls during the higher traffic condition. Non-priority scheme cannot guarantee the QoS level in terms of call blocking probability for the higher priority users during higher traffic condition. Figs. 7 and 9 show that the proposed scheme provides lower call blocking probability for all the traffic classes during the light load traffic condition. However, when the call arrival rate is increased, the proposed scheme only blocks some more calls of the lower priority traffic classes to reduce the call blocking probability of the higher priority traffic classes. Figs. 8 and 10 illustrate that the proposed scheme does not reduce the channel utilization significantly. Fig. 7 shows the call blocking probability analysis when we assume the ratio of call arrival rates of all the four traffic classes are equal. The proposed scheme provides the guaranteed QoS in terms of call blocking probability for the higher priority users. Fig. 8 shows that the proposed scheme also performs well in terms of channel utilization. Our proposed scheme reserved dynamic numbers of channels for different classes of traffic based on the call arrival rates and priority of traffic classes. Whenever the call arrival rate for the higher priority traffic calls is increased, the number of reserved channels for the higher priority traffic calls is also increased in our proposed scheme to reduce the call blocking probability of higher priority traffic calls. Fig. 9 shows that the proposed scheme is able to reduce the call blocking probability of the higher priority traffic class users within a reasonable range even if the call arrival rate of the higher priority traffic class is very high compared to the lower classes of traffic calls. Fig. 10 show that more number of reserved channels to accommodate higher priority traffic calls does not reduce the channel utilization significantly. Whenever the call arrival rate of higher class traffic is very low compared to the lower classes of traffic calls, the proposed scheme reserved less number of channels for the higher priority calls to maintain better channel utilization.

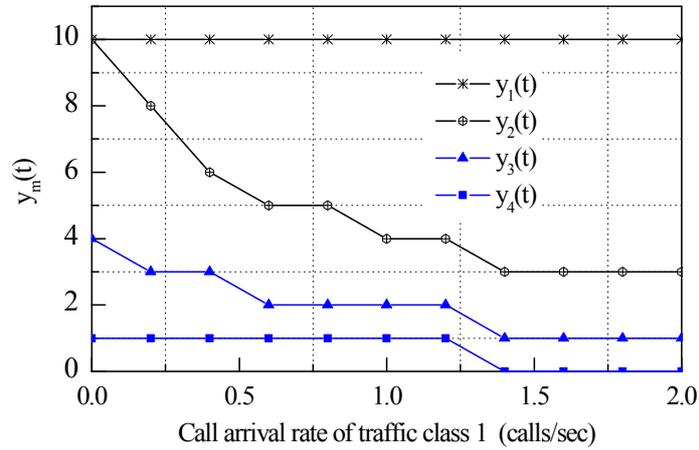

**Fig. 11.** Number of accessible channels among $\Gamma$, for the *m-th* class traffic calls for the condition $\lambda_2$ =0.4 calls/sec, $\lambda_3$ =0.2 calls/sec, and $\lambda_4$ =0.1 calls/sec.

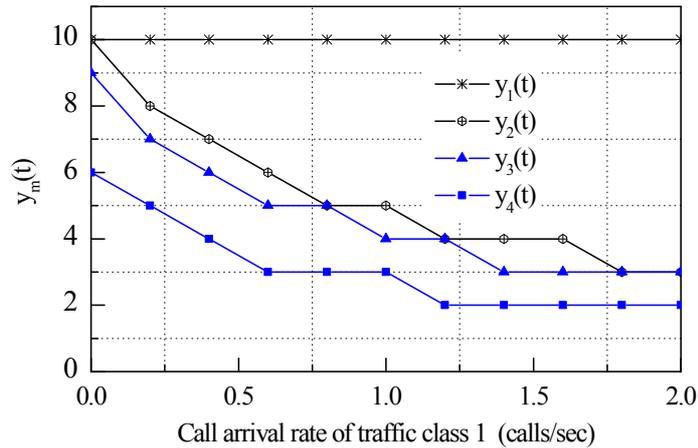

**Fig. 12.** Number of accessible channels among $\Gamma$, for the *m-th* class traffic calls for the condition $\lambda_2$ =0.1 calls/sec, $\lambda_3$ =0.3 calls/sec, and $\lambda_4$ =0.5 calls/sec.

As mentioned before, our proposed scheme reserved a non-fixed numbers of channels for different traffic classes. Figs. 11 and 12 indicate the sharing of $\Gamma$ (=10) channels among different traffic classes for different traffic conditions. The number of accessible channels for each of the traffic classes is varied with the variation of traffic arrival rates to maintain lower call blocking probability of higher priority users and to retain the better channel utilization.

## 5. Conclusions

In this paper, we proposed a dynamic channel allocation scheme for multi-class services in visible light communication system. The proposed scheme can also be successfully applied to other communication systems where multiple traffic classes are provided and resources are allocated based on the priority of the traffic classes. The idea behind the proposed scheme is to reserve a dynamic number of channels for the higher priority users. Reserving channels is equivalent to the guard channels, however the numbers of reserved channels are not fixed in our proposed scheme to maintain higher channel utilization and to provide always lower call blocking probability for the higher priority users. More channels are reserved for the higher priority traffic class when the call arrival rate of higher priority traffic class is higher compared to the lower priority users to support large number of higher priority users. Thus, the scheme gives higher priority for higher priority calls over the lower priority calls without sacrificing the channel utilization.

We have shown that the proposed scheme is quite effective in reducing the call blocking probability of higher priority users without sacrificing the channel utilization. While the proposed scheme blocks some more calls of lower priority calls instead of blocking of higher priority calls during heavy traffic condition. The proposed Markov Chain model will be very much effective for the queuing analysis especially for the priority scheme of any number of traffic classes. The proposed scheme is expected to be of considerable interest for future multi-service VLC networks as well as other wireless networks, since the number of new traffic types with different QoS requirements is expected to further increase with the introduction of new applications.

**Acknowledgements:** This work was supported by the IT R&D program of MKE/KEIT. [10035362, Development of Home Network Technology based on LED-ID]

# Biographies

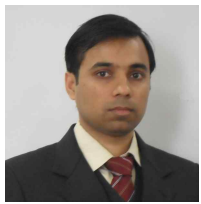

**Mostafa Zaman Chowdhury** received his B.Sc. in Electrical and Electronic Engineering from Khulna University of Engineering and Technology (KUET), Bangladesh in 2002. In 2003, he joined the Electrical and Electronic Engineering department of KUET as a faculty member. He received his M.Sc. in Electronics Engineering from Kookmin University, Korea in 2008. Currently he is working towards his Ph.D. degree in the department of Electronics Engineering at the Kookmin University, Korea. In 2008, he received the excellent student award from the Kookmin University. He served as a reviewer for several international journals (including IEEE Communications Magazine, IEEE Communications Letters, Wireless Networks (springer), and Recent Patents on Computer Science) and IEEE conferences. He has been involved in several Korean Government projects. His research interests include convergence networks, QoS provisioning, mobility management, femtocell networks, and VLC networks.

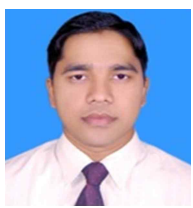

**Muhammad Shahin Uddin** received B.Sc. in Electrical and Electronic Engineering from Rajshahi University of Engineering and Technology (RUET), Bangladesh in 2003. Then, he joined as a faculty member in American International University Bangladesh (AIUB) and then Chittagong University of Engineering and Technology (CUET), Bangladesh. Since 2006 he is with the department of ICT, Mawlana Bhashani Science and Technology University (MBSTU), Bangladesh. He received his M.Sc. in Electronics Engineering from Kookmin University, Korea in 2008. His current research interests focus on VLC networks, LED-ID, sensor network, and wireless cellular networks.

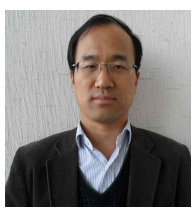

**Yeong Min Jang** received the B.E. and M.E. degree in Electronics Engineering from Kyungpook National University, Korea, in 1985 and 1987, respectively. He received the doctoral degree in Computer Science from the University of Massachusetts, USA, in 1999. He worked for ETRI between 1987 and 2000. Since 2002, he is with the School of Electrical Engineering, Kookmin University, Seoul, Korea. He has organized several conferences such as ICUFN2009 and ICUFN2010. He is currently a member of the IEEE and KICS (Korea Information and Communications Society). He received the Young Science Award from the Korean Government (2003 to 2006). He had been the director of the Ubiquitous IT Convergence Center at Kookmin University since 2005. He has served as the executive director of KICS since 2006. He has served as a founding chair of the KICS Technical Committee on Communication Networks in 2007 and 2008. His research interests include IMT-advanced, RRM, femtocell networks, Multi-screen convergence networks, and VLC WPANs.